\documentstyle[amssymb,twocolumn,aps]{revtex}

\begin{document}
\draft
\title{Magnetization reversal in a ``quasi'' single domain magnetic grain: a new
numerical micromagnetic technique}
\author{Vladimir L. Safonov and H. Neal Bertram}
\address{Center for Magnetic Recording Research, University of California \\
-- San Diego, \\
9500 Gilman Dr., La Jolla, CA 92093-0401, U.S.A.}
\date{\today}
\maketitle

\begin{abstract}
Magnetization reversal in a fine ferromagnetic grain is simulated for the
case of an instantaneously applied reversal magnetic field. The Hamiltonian
of the system contains the exchange interaction, the uniaxial anisotropy,
the Zeeman energy and the dipole-dipole interactions. A cubic grain is
discretized into 64 cubic subgrains and the coupled gyromagnetic equations
of motion are solved without phenomenological damping. A new scheme to solve
these equations is developed that utilizes only two variables per sub-cube
magnetization and strictly conserves the absolute magnitude. The initial
stage of reversal is uniform rotation followed by a nonlinear excitation of
nonuniform magnetic oscillations driven by this uniform mode. An excess of
the initial Zeeman energy is transformed into nonlinear spin waves, allowing
the average magnetization to substantially reverse. The process of
magnetization reversal in fine quasi-single-domain grain exhibits general
features of Hamiltonian wave systems with nonlinear diffusion. This
nonlinear diffusion is forbidden for either a strong reversal field and/or a
small grain size.
\end{abstract}


\section{Introduction}

The problem of magnetization reversal in a fine single-domain grain is of
great importance in magnetic recording physics. Characteristic reversal
times under strong reversal magnetic field give a physical limitation to
both data rates and system signal to noise ratio. Dynamic magnetization
reversal depends on the system or magnetic energy relaxation mechanisms. In
addition to energy absorption to the ``thermal bath'' of the host lattice,
magnetization reversal can occur by excitation of nonlinear `spin-wave' like
modes. In the present paper we explore in detail the excitation of these
modes in a quasi-single-domain particle. This magnetization reversal process
exhibits features of several problems of nonlinear and solid state physics,
such as kinetics of an orientational phase transition and energy transfer in
nonlinear systems of oscillators. Therefore the problem is of interest to
fundamental science.

Dynamic magnetization reversal begins when the external magnetic field
reaches and exceeds the so-called `nucleation' field and the original
remanent state becomes unstable. Typically these dynamical processes in both
single and multidomain magnetic systems are solved by integration of the
Landau-Lifshitz equations \cite{landau} with phenomenological damping \cite
{bertramzhu}. This form of magnetization dynamics describes a coherent
rotation of each discretized subgrain, in which the magnetization magnitude
is conserved. Such approximation is fair for small excitations, as in linear
ferromagnetic resonance {\cite{callen,sparks,gurevich}}, or for some
particular model cases when an excess of energy directly goes to the thermal
bath without excitation of spin waves.

Recently we have explored magnetization reversal in quasi-single-domain
grain by neglecting phenomenological damping and specifically exploring
nonlinear spin-wave like magnetic excitations \cite{safbert},\cite{safbert1}%
. By quasi-single-domain we mean grains whose dimension is on the order of
the domain wall width or in some cases the exchange length. Similar analysis
has also been done on reversal in thin films \cite{boerner},\cite{zhufilm}.
In all these cases, according to Suhl's speculation \cite{suhl2}, the
switching process should be rapid, governed by nonlinear spin-wave
excitations. Recently spin-waves excitated by large magnetization rotations
have been observed experimentally \cite{kabos}.

By neglecting phenomenological damping, the total system magnetic energy is
conserved. These simulations have demonstrated that the excess of Zeeman
energy is transformed to the exchange, anisotropy and magnetostatic energies
by nonlinear spin-wave excitations. The purpose of this paper is to present
a detailed study of magnetization reversal simulations focusing on magnetic
interactions in a single grain.

In addition, we describe here a new numerical scheme of solving the
Landau-Lifshitz equations. Typically numerical micromagnetic simulations
invoke the three Cartesian coordinates of each discretization cube. At each
integration time step the magnetization is modified to satisfy the condition
of constant magnitude. This scheme is preferable to utilization of the
natural two-parameter polar coordinate scheme because of singularities. In
polar coordinates there are always two singularities at the poles. In this
new scheme we utilize two coordinates per magnetization sub-cube, however
here these coordinates are similar to linear combinations of the transverse
magnetization components. Using a simple scheme, all singularieties are
avoided.

In section 2 we discuss the micromagnetic model for our quasi-single-domain
grain. In section 3 we describe the implementation of our new numerical
model. In sections 4, 5 results of simulations are considered in detail.
Then we discuss obtained results and draw conclusions.

\section{Model}

The magnetic energy (Hamiltonian) of the grain can be written in the form: 
\begin{equation}
{\cal H}=\int ({\cal U}_{{\rm exch}}+{\cal U}_{{\rm anis}}+{\cal U}_{{\rm Z}%
}+{\cal U}_{{\rm dmag}})d{\bf r}.  \label{energy}
\end{equation}
Here 
\begin{equation}
{\cal U}_{{\rm exch}}=\frac{A}{M_{{\rm s}}^{2}}\left[ \left( \frac{\partial 
{\bf M}}{\partial x}\right) ^{2}+\left( \frac{\partial {\bf M}}{\partial y}%
\right) ^{2}+\left( \frac{\partial {\bf M}}{\partial z}\right) ^{2}\right]
\label{en-exch}
\end{equation}
is the exchange energy density. 
\begin{equation}
{\cal U}_{{\rm anis}}=K_{{\rm u}}\left[ 1-\left( \frac{{M_{z}}}{M_{{\rm s}}}%
\right) ^{2}\right]  \label{en-anis}
\end{equation}
is the uniaxial anisotropy energy density, 
\begin{equation}
{\cal U}_{{\rm Z}}=-{\bf H}\cdot {\bf M}  \label{en-Z}
\end{equation}
is the Zeeman energy density and ${\cal U}_{{\rm dmag}}$ is the
magnetostatic energy density.

For simplicity we shall consider a micromagnetic grain as a system of $%
n=4\times 4\times 4$ (see, Fig.1) coupled cubic subgrains of the volume $%
V=L^{3}$, where $L$ is the linear size of sub-cube. Each sub-cube can be
characterized as a classical spin:

\begin{equation}
{\bf S}_{j}\equiv -{\bf M}_{j}V/\hbar \gamma ,  \label{classspin}
\end{equation}
where ${\bf M}_{j}$ is the vector magnetization of $j-$th sub-cube, $\hbar $
is Plank's constant and $\gamma >0$ is the gyromagnetic ratio. $|{\bf M}%
_{j}|=M_{{\rm s}}$, where $M_{{\rm s}}$ is the saturation magnetization, and
therefore, $|{\bf S}_{j}|=S\equiv M_{{\rm s}}V/\hbar \gamma $. Each spin is
localized in the center of its sub-cube. The characterization of each
sub-cube magnetization in terms of an effective spin is our choice: the
problem we consider is purely classical. However the spin notation is
convenient for the numerical scheme presented in Sec.3. Spin
characterization also provides a bridge to quantum spin models.

The Hamiltonian Eq.\ (\ref{energy}) of the system in terms of spin notation
can be written as: 
\begin{equation}
{\cal H}={\cal H}_{{\rm exch}}+{\cal H}_{{\rm anis}}+{\cal H}_{{\rm Z}}+%
{\cal H}_{{\rm dmag}}.  \label{hamiltonian}
\end{equation}
Here 
\begin{equation}
{\cal H}_{{\rm exch}}=-\,{\frac{J}{2}}\,{\sum_{j\neq j\prime }}^{\prime }(%
{\bf S}_{j}\cdot {\bf S}_{j\prime }-S^{2})  \label{ham_exch}
\end{equation}
describes the exchange interaction between the nearest neighbors, $%
\,J=2AL/S^{2}$. Note that this form is identical to the usual micromagnetic
approximation that assumes a linear variation of the magnetization between
grain centers \cite{bertramzhu}. The term $-S^{2}$ in (\ref{ham_exch})
yields ${\cal H}_{{\rm exch}}=0$ in the case when all spins are oriented in
the same direction.

\begin{equation}
{\cal H}_{{\rm anis}}=\frac{\hbar \gamma H_{{\rm K}}S}{2}\,\sum_{j=1}^{N}\,%
\left[ 1-\left( \frac{S_{j}^{z}}{S}\right) ^{2}\right]  \label{ham_anis}
\end{equation}
is the uniaxial anisotropy energy, where $H_{{\rm K}}=2K_{{\rm u}}/M_{{\rm s}%
}$ is the anisotropy field. ${\cal H}_{{\rm anis}}=0$ corresponds to all
spins parallel to the anisotropy axis. 
\begin{equation}
{\cal H}_{{\rm Z}}=\hbar \gamma {\bf H}\cdot \sum_{j=1}^{N}{\bf S}_{j}-\hbar
\gamma HNS  \label{ham_Z}
\end{equation}
is the Zeeman energy. ${\cal H}_{{\rm Z}}=0$ if either $H=0$ or all spins
are oriented along the external magnetic field direction.

The magnetostatic interaction of the sub-cubes will be approximated by the
dipole-dipole interaction energy

\begin{eqnarray}
{\cal H}_{{\rm dmag}} &=&{\frac{(\hbar \gamma )^{2}}{2}}\,\sum_{j\neq
j\prime }\left[ \frac{{\bf S}_{j}\cdot {\bf S}_{j\prime }}{r_{jj\prime }^{3}}%
-\frac{3({\bf S}_{j}\cdot {\bf r}_{jj\prime })({\bf S}_{j\prime }\cdot {\bf r%
}_{jj\prime })}{r_{jj\prime }^{5}}\right]  \nonumber \\
&&-{\cal H}_{{\rm dmag}}^{(0)}  \label{ham_dd}
\end{eqnarray}
where ${\bf r}_{jj\prime }={\bf r}_{j}-{\bf r}_{j\prime }$ and ${\bf r}_{j}$%
, ${\bf r}_{j\prime }$ are the radius-vectors of the $j$ and $j^{\prime }$
spins, respectively. The constant ${\cal H}_{{\rm dmag}}^{(0)}$ is chosen so
that ${\cal H}_{{\rm dmag}}=0$ if all spins are oriented upward along the $z$
direction.

We shall study the problem of time evolution of the normalized total
magnetization 
\begin{equation}
{\bf m}=\frac{1}{NM_{{\rm s}}}\,\sum_{j}{\bf M}_{j}=-\frac{1}{NS}\,\sum_{j}%
{\bf S}_{j}  \label{magnetization}
\end{equation}
in the system (\ref{hamiltonian}) from the initial state when $H=0$ and the
averaged magnetization at $t<0$ is oriented ``upward'' in the $z$ direction.
At the moment $t=0$, a strong reversal magnetic field ${\bf H}%
=(H^{x},0,H^{z})$ is applied. The component $H^{z}$ is negative and the
transverse component $H^{x}$ is taken to be relatively small ($|H^{x}|\ll
|H^{z}|\sim H_{{\rm K}}$). The net vector field is sufficient to give only
one energy minimum corresponding to an almost ``downward'' orientation of
the averaged magnetization. Thus the system acquires an excess of Zeeman
energy which can be later transformed to nonlinear spin excitations
containing magnetic anisotropy, exchange and dipole-dipole interaction
energies. This transformation will cause the net magnetization magnitude $|%
{\bf m}|$ to reduce. Mathematically the question of energy transfer is
described in Appendix B.

\section{Numerical scheme}

The Landau-Lifshitz equation without damping in terms of classical spin (\ref
{classspin})\ can be written as 
\begin{eqnarray}
\frac{d{\bf S}_{j}}{dt} &=&-{\bf S}_{j}\times \gamma {\bf H}_{{\rm eff,}j%
{\rm \ }},  \label{merm-ll} \\
\gamma {\bf H}_{{\rm eff,}j} &\equiv &\frac{\partial {\cal H}/\hbar }{%
\partial {\bf S}_{j}}.  \label{efffield}
\end{eqnarray}
This equation describe the evolution of coupled spin components $%
S_{j}^{x},~S_{j}^{y}$ and $S_{j}^{z}$. As time evolves, Eqs.(\ref{merm-ll}),(%
\ref{efffield}) give two system conservations: the net spin of each subgrain
is fixed

\begin{equation}
(S_{j}^{x})^{2}+(S_{j}^{y})^{2}+(S_{j}^{z})^{2}=S^{2},  \label{spin2}
\end{equation}
and the total magnetic energy remains constant. Thus only two independent
variables exist for each $j$-th site.

It is natural to write down only two differential equations for each spin in
terms of spin components or other generalized variables. The most popular
form of ``two-variable'' Landau-Lifshitz equation is known to be in
spherical coordinates. However the use of spherical coordinates for an
arbitrary motion of spin gives a stiff system of differential equations with
singularities at the poles. Moreover, numerical calculations of
trigonometric functions typically take much longer time than simple
arithmetic operations. This is why the spherical coordinate approach is not
usually used in micromagnetic calculations. In practice the direct method of
numerical solution of the Landau-Lifshitz equation is used (see, for
example, \cite{Schabes,nakatani}). Dynamic equations for all three variables
are integrated with continuum numerical modification so that the Eq. (\ref
{spin2}) is maintained.

In this paper we present a new scheme for numerical solution of the
Landau-Lifshitz equation (\ref{merm-ll}), in which only two variables are
utilized and condition (\ref{spin2}) is automatically maintained. We
introduce the real and dimensionless variables $q_{j}$ (as a generalized
coordinate)\ and $p_{j}$ (as a generalized momentum) as follows 
\begin{eqnarray}
S_{j}^{x} &=&q_{j}\,\sqrt{(S+S_{j}^{z})/2},  \nonumber \\
S_{j}^{y} &=&p_{j}\,\sqrt{(S+S_{j}^{z})/2},  \nonumber \\
S_{j}^{z} &=&S-(q{_{j}^{2}+}p{_{j}^{2}})/2.  \label{hopri-sch}
\end{eqnarray}
These transformations are exact and can be considered to be a form of
classical Holstein-Primakoff \cite{hopri} transformation (see Appendix).
Schl\"{o}mann \cite{schlo} also presented a classical spin in a similar form.

The equations of motion for $q_{j}$ and $p_{j}$ are the exact analog of the
Landau-Lifshitz equations (\ref{merm-ll}) and can be written in the form

\begin{mathletters}
\begin{eqnarray}
\frac{dq_{j}}{d\tau } &=&\frac{{\bf H}_{{\rm eff,}j}}{H_{{\rm K}}}\cdot 
\frac{\partial {\bf S}_{j}}{\partial p_{j}},  \label{q1-LL} \\
\frac{dp_{j}}{d\tau } &=&-\frac{{\bf H}_{{\rm eff,}j}}{H_{{\rm K}}}\cdot 
\frac{\partial {\bf S}_{j}}{\partial q_{j}},  \label{p1-LL}
\end{eqnarray}
where a dimensionless time $\tau =\gamma H_{{\rm K}}t$ is introduced.
Corresponding spin derivatives are

\end{mathletters}
\begin{eqnarray}
\frac{\partial S_{j}^{x}}{\partial q_{j}} &=&\sqrt{\frac{S+S_{j}^{z}}{2}}-%
\frac{q_{j}^{2}}{4}\sqrt{\frac{2}{S+S_{j}^{z}}},  \nonumber \\
\frac{\partial S_{j}^{y}}{\partial p_{j}} &=&\sqrt{\frac{S+S_{j}^{z}}{2}}-%
\frac{p_{j}^{2}}{4}\sqrt{\frac{2}{S+S_{j}^{z}}},  \nonumber \\
\frac{\partial S_{j}^{x}}{\partial p_{j}} &=&\frac{\partial S_{j}^{y}}{%
\partial q_{j}}=-\frac{q_{j}p_{j}}{4}\sqrt{\frac{2}{S+S_{j}^{z}}},  \nonumber
\\
\frac{\partial S_{j}^{z}}{\partial q_{j}} &=&-q_{j},\quad \frac{\partial
S_{j}^{z}}{\partial p_{j}}=-p_{j}.  \label{sderiv1}
\end{eqnarray}

It is necessary to mention that Eqs.(\ref{q1-LL}), (\ref{p1-LL}) contain a
singularity at $S+S_{j}^{z}=0$. To avoid this singularity we shall use (\ref
{q1-LL}), (\ref{p1-LL}) only in the hemisphere $|{\bf S}_{j}|=S$\ with $%
S_{j}^{z}\geq 0$. This is illustrated in Fig.2. Following Eq.(5), negative
magnetization corresponds to a positive spin and the ($p,q$) coordinates are
in the upper half hemisphere. Initially before reversal, when $M_{z}>0$ it
is convenient to introduce analogous variables $Q_{j}$ and $P_{j}$ (see
Appendix A) for the hemisphere $|{\bf S}_{j}|=S$\ with $S_{j}^{z}<0$ :

\begin{eqnarray}
S_{j}^{x} &=&Q_{j}\,\sqrt{(S-S_{j}^{z})/2},  \nonumber \\
S_{j}^{y} &=&-P_{j}\,\sqrt{(S-S_{j}^{z})/2},  \nonumber \\
S_{j}^{z} &=&-S+(Q{_{j}^{2}+}P{_{j}^{2}})/2.  \label{hp2}
\end{eqnarray}
The equations of motion for $Q_{j}$ and $P_{j}$ have the form:

\begin{mathletters}
\begin{eqnarray}
\frac{dQ_{j}}{d\tau } &=&\frac{{\bf H}_{{\rm eff,}j}}{H_{{\rm K}}}\cdot 
\frac{\partial {\bf S}_{j}}{\partial P_{j}},  \label{q2-LL} \\
\frac{dP_{j}}{d\tau } &=&=-\frac{{\bf H}_{{\rm eff,}j}}{H_{{\rm K}}}\cdot 
\frac{\partial {\bf S}_{j}}{\partial Q_{j}},  \label{p2-LL}
\end{eqnarray}
and spin derivatives are

\end{mathletters}
\begin{eqnarray}
\frac{\partial S_{j}^{x}}{\partial Q_{j}} &=&\sqrt{\frac{S-S_{j}^{z}}{2}}-%
\frac{Q_{j}^{2}}{4}\sqrt{\frac{2}{S-S_{j}^{z}}},  \nonumber \\
\frac{\partial S_{j}^{y}}{\partial P_{j}} &=&-\sqrt{\frac{S-S_{j}^{z}}{2}}+%
\frac{P_{j}^{2}}{4}\sqrt{\frac{2}{S-S_{j}^{z}}},  \nonumber \\
\frac{\partial S_{j}^{x}}{\partial P_{j}} &=&-\frac{\partial S_{j}^{y}}{%
\partial Q_{j}}=-\frac{Q_{j}P_{j}}{4}\sqrt{\frac{2}{S-S_{j}^{z}}},  \nonumber
\\
\frac{\partial S_{j}^{z}}{\partial Q_{j}} &=&Q_{j},\quad \frac{\partial
S_{j}^{z}}{\partial P_{j}}=P_{j}.  \label{sderiv2}
\end{eqnarray}

A continuous transition from one set of variable to another one is defined by

\begin{eqnarray}
q_{j}\, &=&Q_{j}\,\sqrt{\frac{S-S_{j}^{z}}{S+S_{j}^{z}}}{,}  \nonumber \\
p_{j}\, &=&-P_{j}\,\sqrt{\frac{S-S_{j}^{z}}{S+S_{j}^{z}}}{,}  \label{qp-QP}
\end{eqnarray}

and

\begin{eqnarray}
Q_{j} &=&q_{j}\,\sqrt{\frac{S+S_{j}^{z}}{S-S_{j}^{z}}}{,}  \nonumber \\
P_{j} &=&-p_{j}\,\sqrt{\frac{S+S_{j}^{z}}{S-S_{j}^{z}}}{.}  \label{QPqp}
\end{eqnarray}

Thus, we begin the calculations with $P_{j}(0)$ and$\ Q_{j}(0)$. After a
short integration time $\Delta t$\ we find $P_{j}(\Delta t)$ and$\
Q_{j}(\Delta t)$ and calculate corresponding $S_{j}^{x}$, $S_{j}^{y}$ and $%
S_{j}^{z}$. If $S_{j}^{z}<0$, we continue integration of the Landau-Lifshitz
equations with $P_{j}$ and$\ Q_{j}$. In the case when $S_{j}^{z}\geq 0$, the 
$j-$th spin trajectory crosses the equator and enters the upper hemisphere.
To avoid the singularity in the $P_{j}$ and$\ Q_{j}$ coordinates in the
upper pole (see Fig.2), the new variables $p_{j}$ and $q_{j}$ are introduced
by Eq.(\ref{qp-QP}). Integration of the Landau-Lifshitz equations using $%
p_{j}$ and $q_{j}$ until the $z$-component of the spin changes sign ($%
S_{j}^{z}<0$). At that point we change back to the $P_{j}$ and$\ Q_{j}$
coordinates.

It should be noted that the Eqs.(\ref{q1-LL}), (\ref{p1-LL}), (\ref{q2-LL})
and (\ref{p2-LL}) have simple algebraic operations only and therefore
numerical integration of these equations do not contain time consuming
operations.

A characteristic time scale of the system is $1/\gamma H_{{\rm K}}$. The
role of exchange interaction is defined by the parameter $H_{{\rm e}}/H_{%
{\rm K}}$, where $\,H_{{\rm e}}=JS/\hbar \gamma =2A/M_{{\rm s}}L^{2}\,$ is
the exchange field from the nearest neighbor. One can write

\begin{equation}
H_{{\rm e}}/H_{{\rm K}}=(n^{2/3}/16)(\delta _{{\rm w}}/\delta _{{\rm s}%
})^{2},  \label{wall_length}
\end{equation}
where $\delta _{{\rm w}}=4(A/K_{{\rm u}})^{1/2}$ is the domain wall width, $%
\delta _{{\rm s}}=n^{1/3}L=4L$ is the linear sample size. The exchange
interaction energy can also be characterized by, so-called, ``exchange
length'' $\,\delta _{{\rm ex}}=A^{1/2}/M_{s}\,$. Therefore one can write 
\begin{equation}
H_{{\rm e}}/H_{{\rm K}}=2n^{2/3}(M_{{\rm s}}/H_{{\rm K}})(\delta _{{\rm ex}%
}/\delta _{{\rm s}})^{2}.  \label{exch_length}
\end{equation}

We shall focus on the case of relatively strong uniaxial anisotropy when $M_{%
{\rm s}}/H_{{\rm K}}\ll 1$ (which is valid for real magnetic recording
grains). In the zeroth order of this small parameter we can neglect the
dipole-dipole interaction in the system. However as it will be demonstrated
in the next section that even relatively small dipole-dipole interaction,
which absorbs very small amount of the Zeeman energy plays an important role
in the magnetization reversal.

\subsection{Numerical implementation and initial conditions}

The system of Landau-Lifshitz equations has been numerically solved for $%
n=64 $. High-accuracy solutions to these ordinary differential equations
were obtained with minimal computational effort by utilizing the
Bulirsch-Stoer method with adaptive stepsize control \cite{numrec}.
Integration steps\ were about four orders less than the characteristic time $%
1/\gamma H_{{\rm K}}$.

The equations of motion (\ref{q1-LL}), (\ref{p1-LL}), (\ref{q2-LL}), (\ref
{p2-LL}) contain relative parameters only. For simulations these parameters
were chosen to correspond to real magnetic recording media. For example,
with $A=10^{-6}$ erg/cm, $H_{{\rm K}}=7$ kOe and $M_{{\rm s}}=250$ Oe, the
scaled parameters are $\delta _{{\rm w}}\simeq 43$ nm, $\delta _{{\rm ex}%
}=40 $ nm and $M_{{\rm s}}/H_{{\rm K}}=0.036$ .

Nonlinearities can only be excited were there are deviations in the
magnetization state from perfect alignment. We include this by allowing the
system to initially be in thermal equilibrium. We assign randomly a total
energy deviation of $64k_{{\rm B}}T$. Practically we randomly deviate the
magnetization of each cell within a fixed limit. This limit is adjusted to
achieve the above total energy deviation. To obtain a true thermal
equilibrium the Landau-Lifshitz equations are solved, given the above
initialization, for a time $\gamma H_{{\rm K}}t_{0}\sim 200$. At this point
the system is prepared for the application of the reversal field.

\section{Magnetization reversal without dipole interactions}

First we shall consider the case when the dipole-dipole interactions in $%
\gamma {\bf H}_{{\rm eff,}j}$ (\ref{efffield}) are neglected. Fig.3a,b show
the time dependencies of $m_{z}(t)$ and $|{\bf m}(t)|$ in the case when the
reverse magnetic field has a very small deviation from the anisotropy axis ($%
H_{x}/H_{z}=0.02$). Fig.3b is simply an extension of the time axis of
Fig.3a. The field magnitude is slightly less than the nucleation field for
uniform rotation ($H_{nuc}\simeq -0.9H_{{\rm K}}$). This field corresponds
(by numerical test) to the lowest reverse field required to produce
nonlinear excitations. The sample dimension is 16\% and 25\% greater than
the domain wall width and exchange length, respectively. Initially the
longitudinal component of magnetization decreases monotonically from $%
m_{z}\simeq 1$ to $\simeq 0.7$ and then at a `critical' time $\gamma H_{{\rm %
K}}t_{c}\simeq 70$, oscillations of $m_{z}$ occur. In contrast, the net
magnitude of magnetization remains practically constant $|{\bf m}|\simeq 1$\
until $t_{c}$ and then abruptly decreases exhibiting irregular oscillations.
For times less than $t_{c}$ the average magnetization simply precesses about
the reverse field. For times greater $t_{c}$ the excitation of nonlinear
spin waves becomes pronounced. These excitations both absorb Zeeman energy
and reduce the magnetization.

This behavior is typical of nonlinear spin-wave processes whose amplitudes
grow exponentially from an initial thermal level. The decrease in net
magnetization becomes observable at virtually a critical time $t_{c}$\ when
the magnetic moment of the (transverse) excited waves becomes comparable
with the net magnetic moment. At the same time the system nonlinearities
restrict the exponential growth. The oscillations in $m_{z}$ that occur for $%
t>t_{c}$ primarily correspond to the uniform rotation: the frequency
increases with decreasing $m_{z}$. If spin-wave excitations did not occur,
oscillations in $m_{x}$, $m_{y}$, $m_{z}$ would reflect uniform rotation
only and at every instant sum to give constant $|{\bf m}|\simeq 1$.

With increasing time the average $m_{z}$ continually decreases reaching, for
this example an asymptotic level in the vicinity of $m_{z}=0$ (at least for
times less than $\gamma H_{{\rm K}}t\sim 3000$). There is no reversal of $%
m_{z}(t)$. The average magnetization $|{\bf m}(t)|$ also decreases by about $%
\approx 60\%$. As time proceeds, the spin-wave oscillations exhibit
increased chaotic behavior.

Fig.4 shows similar time dependences as in Fig.3 but for the case of
increased angle of the reverse magnetic field from the anisotropy axis ($%
H_{x}/H_{z}=0.10$). The field magnitude ($H_{z}/H_{{\rm K}}=-0.70$) also
corresponds to the onset of spin-wave excitation. As expected, the
magnetization dynamics is much faster in this case. We can see the beginning
of nonlinear spin-wave excitation at $\gamma H_{{\rm K}}t_{c}\simeq 30$,
where the $m_{z}$ has reduced due to pure rotation to $\simeq 0.3$. The
decrease of longitudinal magnetization is more substantial, passing through
zero at $\gamma H_{{\rm K}}t_{0}\simeq 75$ and reaching about $\approx 55\%$
of its nominal value in the opposite direction after $\gamma H_{{\rm K}%
}t\gtrsim 1000$. Similar to the case shown in Fig.3, the average
magnetization $|{\bf m}(t)|$ decreases to $\approx 60\%$. Note that the
amplitudes of chaotic oscillations of $|{\bf m}(t)|$\ are relatively smaller
than in the case of Fig.3 (compare the long time behavior of Fig.3b and
Fig.4b). Here most of the Zeeman energy is transformed into high frequency
exchange-dominant spin-wave modes, which have relatively small magnetic
moments.

As discussed earlier, the total magnetic energy remains constant in these
dynamic processes. In Fig.5a,b we plot the evolution of the Zeeman, exchange
and anisotropy energies (Fig.5b is an extended scale of Fig.5a). The reverse
angle is intermediate between Fig.3 and 4. Initially, for times less than $%
t_{c}$ (in this case $\gamma H_{{\rm K}}t_{c}\simeq 45$), uniform rotation
yields a decrease of the Zeeman energy and an increase of the anisotropy
energy. At the onset of nonlinear spin-wave excitation we see, in addition,
an increase of the exchange energy. The wave-like nature of these
excitations can be seen in Fig.5a in the beating of the exchange and
anisotropy energies. The nonlinear excitation of chaotic oscillations of the
magnetization can be thought of as an ``overheating of the magnetic
system''. The more exchange magnetic modes that are involved in the process,
the smaller the averaged magnetization $|{\bf m}(t)|$.

Fig.6a-d demonstrate the time dependencies of $m_{z}(t)$ and $|{\bf m}(t)|$
for various magnetic fields: $-0.78H_{{\rm K}}$, $-0.79H_{{\rm K}}$, $%
-0.90H_{{\rm K}}$ and $-1.0H_{{\rm K}}$, respectively. As in Fig.5 the field
angle is fixed at $H_{x}/H_{z}=0.05$ and relative sizes are unchanged. For
this angle the Stoner-Wohlfarth nucleation field is $-0.83H_{{\rm K}}$. In
this case the onset field for nonlinear spin-wave excitation is $H_{z}$ =$%
-0.79H_{{\rm K}}$. There is no reversal below this field as shown in Fig.6a.
The magnetization evolution at the onset field is shown in Fig.6b (similar
to Figs.3,4). Magnetization dynamics for larger fields are shown in
Figs.6c,d. A notable features can be seen: the critical time for nonlinear
spin-wave excitation increases with increasing field. The increase of $t_{c}$
allows for several cycles of uniform rotation to occur prior to the
nonlinear spin-wave excitation. No reversal dynamics was observed\ beyond $%
H_{z}<-1.1H_{{\rm K}}$ (in this particular case). Essentially as the reverse
field is increased beyond the spin-wave onset field, $t_{c}$ rapidly
increases and becomes infinite. These phenomena also occur for other field
angles ($H_{x}/H_{z}$).

As indicated in Figs.6a-d the asymptotic, long time value of $m_{z}$
increases with increasing applied field (i.e. less magnetization reversal
occurs). In Fig.7 we explicitly plot the asymptotic values of averaged $%
\langle m_{z}(t)\rangle $ and $\langle |{\bf m}(t)|\rangle $ versus applied
field. It should be noted that the excitation of nonlinear spin waves occurs
only for a narrow band of applied fields.

Fig.8 shows the evolution of magnetization at various sample sizes. We see
that a decrease of the sample size relative to the domain wall width leads
to a suppression of magnetization reversal. This fact is also can be
interpreted in terms of nonlinear spin-wave processes. By decreasing the
sample size, we increase the frequencies of the non-uniform spin-wave modes
and therefore restrict the number of resonance nonlinear processes that can
contribute to a decay of the uniform mode. Analysis for larger sample would
require finer discretization.

\section{Magnetization reversal with dipole interactions}

In this section we show numerical simulations that include the dipole-dipole
interactions in $\gamma {\bf H}_{{\rm eff,}j}$ (\ref{efffield}). The
relative role of dipole-dipole interaction is small because we assume a
small magnetostatic parameter $M_{{\rm s}}/H_{{\rm K}}\ll 1$. In this case
we can not expect a noticeable absorption of the Zeeman energy solely by the
dipole-dipole reservoir. Nevertheless, the dipole-dipole interaction plays
an important role in the process of energy transfer: it reduces the symmetry
of the system and therefore opens new channels of nonlinear spin-wave
interactions.

Let us compare the results of Fig.3 obtained without dipole interaction and
corresponding results in Fig.9 with the dipole interaction. The onset time $%
t_{c}$ of nonlinear excitation is about the same. However $m_{z}(t)$
decreases more rapidly, reverses and reaches an asymptotic value of $\simeq
-0.5$. This is in contrast to Fig.2 where the asymptotic $m_{z}(t)$ does not
reverse. In this case the average magnetization $|{\bf m}(t)|$ decreases to $%
\approx 50\%$, slightly more than the $\approx 60\%$ in Fig.3. The
amplitudes of chaotic oscillations of $|{\bf m}(t)|$ in this case are much
smaller than occur in Fig.3. The excess Zeeman energy is transformed into
exchange modes, which have smaller magnetic moments.

Plotted in Fig.10 is the magnetization dynamics for a larger field angle ($%
H_{x}/H_{z}=0.10$). As compared with Fig.4, including magnetostatic
interactions slightly increases the reversal rate. The longitudinal
magnetization passes through zero at $\gamma H_{{\rm K}}t_{0}\simeq 55$ and,
similar to Fig.4, reaches about $\approx 55\%$ of its nominal value in the
opposite direction after $\gamma H_{{\rm K}}t\gtrsim 500$. The average
magnetization $|{\bf m}(t)|$ is decreased to $\approx 60\%$.

Fig.5c,d demonstrate typical transformations of Zeeman, anisotropy and
exchange energies for short (c) and long (d) time intervals for
magnetization dynamics with dipole interactions. We see faster dynamics of
energy transformations than shown in Fig.5a,b. It should be noted that the
energy absorbed by dipole interactions is relatively small (within $\pm 0.01$
in relative units) as expected and therefore not shown. However the
magnetostatic interactions change the balance the exchange and anisotropy
energies.

In Fig.11 are shown the time dependencies $m_{z}(t)$ and $|{\bf m}(t)|$ for
various reversal fields. In agreement with Fig.6, the reversal dynamics is
suppressed with increasing the strength of reversal field but the range of
reversal dynamics is much broader. We did not observe any reversal dynamics
at $H_{z}\leq -3H_{{\rm K}}$ in this particular case. We believe that the $%
4\times 4\times 4$ discretization is not a major limiting factor. For the
case of $n=1000$ subgrains no reversal dynamics was observed at $H_{z}\leq
-4H_{{\rm K}}$ \cite{safbert1}.

In Fig.12 we plot the asymptotic relative change in Zeeman energy $\Delta E_{%
{\rm Z}}=(1-\langle m_{z}\rangle )(-H_{z}/H_{{\rm K}})$ versus the applied
magnetic field for the parameters of Fig.11 (solid dots). Even though there
is less asymptotic magnetization decrease as the field is increased, the
relative change in Zeeman energy does increase. After the onset of nonlinear
spin-wave excitation, in this case at $H_{z}=-0.79H_{{\rm K}}$, the amount
of spin-wave energy that can be excited increases to a maximum value. In
addition (open dots), the change in Zeeman energy is plotted corresponding
to Fig.5. for the case of no magnetostatic interaction. The same trend
occurs but only over a small interval of field values.

Fig.13 shows the evolution of magnetization at various sample sizes. As in
Fig.8, the decrease of the sample size relative to the domain wall width
leads to a suppression of magnetization reversal. The minimum size for
spin-wave excitation remains the same, however differences are seen in the
amplitude of the chaotic excitations.

We have also examined the effect of grain size relative to the exchange
length at fixed relative domain wall size ($\delta _{{\rm s}}/\delta _{{\rm w%
}}=1.16$). The system evolution is more sensitive to the change of exchange
length if the dipole-dipole interaction is taken into account. The primary
effect is that with a smaller relative exchange length the rate of decrease
of the average magnetization is faster. As discussed above, the presence of
magnetostatic interactions allows the excitation of additional spin-wave
modes that were otherwise degenerate. These modes are more easily excited if
the sample size is large compared to the exchange length.

\section{Discussion}

Magnetization reversal occurs by reducing the Zeeman energy as the
magnetization goes from an initial high energy state to one of low energy
(approximately in the field direction). Typically this dynamic process is
analyzed using the Landau-Lifshitz equation with an appropriate
phenomenological damping term. Here we have excluded phenomenological
damping and shown that substantial, rapid reversal can occur. In this case
the total magnetic energy is conserved, however subject to constraints of
sample size and field range, the Zeeman energy is reduced by the excitation
of nonlinear spin-waves.

In the present study we have specifically focused on a cubic grain
sufficiently small to be almost single domain. In this case after uniform
rotation over a short time period, nonlinear spin waves are excited and the
average grain magnetization decreases. For the sample sizes examined here
the reduction was from positive saturation ($+M_{{\rm s}}$) to at most, $%
-0.6M_{{\rm s}}$. Due to numerical limitations we did not examine large
grains and it is possible that, with sufficient discretization, almost
complete reversal might occur for samples much larger than the domain wall
width. In this study we considered only samples with small relative
magnetostatic energy contribution. Studies have been performed on high
moment low anisotropy thin films, where a large out of plane magnetization
reversal torque occurs \cite{boerner},\cite{zhufilm}. In these cases
virtually complete reversal almost always occurs with no phenomenological
damping.

The decrease and suppression of magnetization reversal with increasing the
reversal field strength initially seem unusual. In the discussion of Fig.11
a limit is suggested as due to the saturation of the spin-wave energy.
However, the problem we consider is a general system of nonlinear
oscillators (nonlinear spin waves). In the oscillator problem one mode (here
corresponding to coherent rotation of the magnetization) is\ initially
strongly excited. One might expect that the system would move towards its
most probable state (thermal equilibrium) by a chaotic mode mixing. With any
large reverse field we would expect reversal to a new equilibrium state.
However, the observed reversal decreases and suppression can be
qualitatively understood as a result of nonlinear spin-wave interactions.

The processes of nonlinear spin-wave decay are defined by several factors.
First, it is necessary that the frequency of the initial mode and some
portion of the nonlinear spin-wave spectrum overlap. Second, we require
sufficiently strong nonlinear interwave coupling. These factors determine a
threshold amplitude of the decaying wave (as an effective pumping field).
Third, the amplitude of initial (decaying) wave must exceed a threshold
value. When the reversal field strength is increased, all these three
requirements eventually are not met. As suggested in Ref.\cite{suhl2},
increasing the reverse field shifts the spin-wave spectrum away from the
uniform precession mode. Also as the reversal field is increased nonlinear
coupling decreases. At a certain reversal field the threshold of nonlinear
decay will exceed the amplitude of coherent magnetization rotation and the
reversal process will be suppressed. Generally speaking, such a suppression
of the energy transfer from the excited mode to other modes is a common
feature of the theory of Hamiltonian systems of nonlinear oscillators (see,
e.g.,\cite{chirikov}). The most simple and popular example of nonlinear
coupled oscillators is the Fermi-Pasta-Ulam chain (see, e.g.,\cite{fpu}).

The numerical scheme we have presented is valuable here primarily because
the number of coordinates is reduced by 2/3. Thus, the computational storage
requirement is correspondingly reduced. However, there is an additional
advantage to this formulation. These coordinates correspond naturally to
nonlinear oscillators that describe spin-wave motions. Recently, the problem
of thermally agitated reversal has been analyzed\ in a micromagnetic grain.
Using this nonlinear oscillator representation a simple exact solution has
been obtained for the first passage time \cite{safonov},\cite{safonbertram},
\cite{bertramsaf}.

\section{Conclusion}

Magnetization reversal in a quasi-single-domain grain can occur by nonlinear
spin-wave excitation. In this process an excess of Zeeman energy is
transfered to uniaxial anisotropy and exchange energies. The dipole-dipole
interaction plays an important role even if small ($M_{{\rm s}}/H_{{\rm K}%
}\ll 1$). The process of magnetization reversal in a fine grain exhibits
general features of system of interacting oscillators: energy transfer by
nonlinear resonance processes. This nonlinear process (diffusion) is
forbidden for either a strong reversal field and/or a small grain size.

We have presented a new scheme to perform numerical integration of the
Landau-Lifshitz equations. This new scheme utilizes only two variables per
discretization cell, but strictly conserves the magnetization magnitude.
Comparison with conventional techniques show that both schemes require
approximately the same computation time at the same level of accuracy (the
new scheme takes a slightly shorter time). The main advantage of the new
scheme is that it utilizes a smaller number of variables (2/3 of the total
magnetization components) and therefore requires less memory.

\section*{ Acknowledgments}

The authors are grateful to Prof. Harry Suhl for valuable comments and
discussions. We also thank E. Boerner, M. E. Schabes and A. N. Slavin for
interesting and helpful discussions of obtained results. This work was
partly supported by matching funds from the Center for Magnetic Recording
Research at the University of California - San Diego and CMRR incorporated
sponsor accounts.

\appendix 

\section{Equations of motions and canonical variables}

\subsection{Classical commutations}

Hamilton's equations can be written in a general form

\begin{equation}
i\hbar \,\frac{d{\cal O}}{dt}=[\![\,{\cal O},{\cal H}\,]\!]  \label{hameq}
\end{equation}
Here ${\cal O}$ is a function of interest and $[\![\,\ldots ,\ldots
\,]\!]=i\{\ldots ,\ldots \}$ denotes a classical analog of commutator and $%
\{\ldots ,\ldots \}$ are the Poisson brackets. Plank's constant $\hbar $ is
used as a dimensional constant to simplify a dimensional structure of the
equation and for simple correspondence with quantum mechanics.

Taking into account the Mermin's \cite{mermin} formula for the Poisson
brackets for a system of classical spins ${\bf S}_{j}$, one can write: 
\begin{equation}
\lbrack \![\,{\cal A},{\cal B}\,]\!]=i\sum_{j}{\bf S}_{j}\cdot \frac{%
\partial {\cal A}}{\partial {\bf S}_{j}}\times \frac{\partial {\cal B}}{%
\partial {\bf S}_{j}}.  \label{merm}
\end{equation}
It is simple to show that the commutation rules for the classical spins are
the same as in quantum mechanics: 
\begin{equation}
\lbrack \![\,S_{j}^{z},S_{j\prime }^{\pm }\,]\!]=\pm S_{j}^{\pm }\delta
_{jj\prime },\quad \lbrack \![\,S_{j}^{+},S_{j\prime
}^{-}\,]\!]=2S_{j}^{z}\delta _{jj\prime }.  \label{spincom}
\end{equation}
where $S_{j}^{\pm }=S_{j}^{x}\pm iS_{j}^{y}$ are the circular spin
components.

Substituting ${\cal O}={\bf S}_{j}$ to Eq.\ (\ref{hameq}) and taking into
account that

\begin{eqnarray*}
\lbrack \![\,{\bf S},{\cal H}\,]\!] &=&i{\bf S}\cdot \frac{\partial {\bf S}}{%
\partial {\bf S}}\times \frac{\partial {\cal H}}{\partial {\bf S}} \\
&=&-i\frac{\partial {\bf S}}{\partial {\bf S}}\cdot {\bf S}\times \frac{%
\partial {\cal H}}{\partial {\bf S}} \\
&=&-i{\bf S}\times \frac{\partial {\cal H}}{\partial {\bf S}},
\end{eqnarray*}
one can obtain the Landau-Lifshitz equation (\ref{merm-ll}).

\subsection{Holstein-Primakoff transformations}

It may be easily shown that the commutations (\ref{spincom}) are valid if 
\begin{eqnarray}
S^{z} &=&S-a^{\ast }a,  \nonumber \\
S^{+} &=&a\,\sqrt{S+S^{z}},  \nonumber \\
S^{-} &=&a^{\ast }\,\sqrt{S+S^{z}}.  \label{holstein-primakoff1}
\end{eqnarray}
These relations describe the classical spin realization in terms of complex
variables $a^{\ast }$ and $a$ corresponding to quantum creation and
annihilation operators, respectively. The transformation (\ref
{holstein-primakoff1}) in the quantum case was introduced by Holstein and
Primakoff \cite{hopri}. Suhl \cite{suhl1}\ first utilized approximate
Holstein-Primakoff transformation with complex variables to describe
magnetic resonance.

The classical analog of the commutator for $\,a^{\ast }\,$ and $\,a\,$ is

\begin{equation}
\lbrack \![{\cal A},{\cal B}]\!]\equiv {\frac{{\partial }{\cal A}}{{\partial 
}a}}{\frac{{\partial }{\cal B}}{{\partial }a^{\ast }}}-{\frac{{\partial }%
{\cal B}}{{\partial }a}}{\frac{{\partial }{\cal A}}{{\partial }a^{\ast }}}.
\label{osccomm}
\end{equation}
The Eq.(\ref{hameq}) with (\ref{osccomm}) can be rewritten as

\begin{equation}
\frac{da}{dt}=-i\frac{\partial {\cal H}/\hbar }{\partial {\bf S}}\cdot \frac{%
\partial {\bf S}}{\partial a^{\ast }}=-i\gamma {\bf H}_{{\rm eff}}\cdot 
\frac{\partial {\bf S}}{\partial a^{\ast }}.  \label{aLL}
\end{equation}

Another form of Holstein-Primakoff transformation can be written as

\begin{eqnarray}
S^{z} &=&-S-b^{\ast }b,  \nonumber \\
S^{+} &=&b^{\ast }\,\sqrt{S-S^{z}},  \nonumber \\
S^{-} &=&b\,\sqrt{S-S^{z}}.  \label{holstein-primakoff2}
\end{eqnarray}
The classical analog of commutator for $\,b^{\ast }\,$ and $\,b\,$ is

\[
\lbrack \![{\cal A},{\cal B}]\!]\equiv {\frac{{\partial }{\cal A}}{{\partial
b}}}{\frac{{\partial }{\cal B}}{{\partial b}^{\ast }}}-{\frac{{\partial }%
{\cal B}}{\partial b}}{\frac{{\partial }{\cal A}}{{\partial b}^{\ast }}} 
\]
and the corresponding equation of motion has the form

\begin{equation}
\frac{db}{dt}=-i\frac{\partial {\cal H}/\hbar }{\partial {\bf S}}\cdot \frac{%
\partial {\bf S}}{\partial b^{\ast }}=-i\gamma {\bf H}_{{\rm eff}}\cdot 
\frac{\partial {\bf S}}{\partial b^{\ast }}.  \label{bLL}
\end{equation}
It should be noted that Eqs.(\ref{aLL}) and (\ref{bLL}) represent other
forms of the Landau-Lifshitz equation (\ref{merm-ll}).

\subsection{Real canonical variables}

Let us introduce real variables $q$ and $p$ as

\[
a=\frac{q+ip}{\sqrt{2}},\qquad a^{\ast }=\frac{q-ip}{\sqrt{2}} 
\]
Thus the Eqs.(\ref{holstein-primakoff1}) become 
\begin{eqnarray}
S^{x} &=&q\,\sqrt{(S+S^{z})/2},  \nonumber \\
S^{y} &=&p\,\sqrt{(S+S^{z})/2},  \nonumber \\
S^{z} &=&S-(q{^{2}+}p{^{2}})/2.  \label{6}
\end{eqnarray}
These formulas are exact and can be considered as a form of classical
Holstein-Primakoff transformation.

The classical commutator for $q$ and $p$ is: 
\begin{equation}
\lbrack \![{\cal A},{\cal B}]\!]\equiv {i}\left( \frac{\partial {\cal A}}{%
\partial q}\frac{\partial {\cal B}}{\partial p}-\frac{\partial {\cal B}}{%
\partial q}\frac{\partial {\cal A}}{\partial p}\right) .  \label{pq-comm}
\end{equation}
The equations of motion for real and dimensionless $q$ and $p$ are 
\begin{eqnarray}
\frac{dq}{dt} &=&\frac{\partial {\cal H}/\hbar }{\partial p}=\frac{\partial 
{\cal H}/\hbar }{\partial {\bf S}}\cdot \frac{\partial {\bf S}}{\partial p},
\nonumber \\
\frac{dp}{dt} &=&-\,\frac{\partial {\cal H}/{\hbar }}{\partial q}=-\frac{%
\partial {\cal H}/\hbar }{\partial {\bf S}}\cdot \frac{\partial {\bf S}}{%
\partial q}.  \label{eq-qp}
\end{eqnarray}
We shall use the transformation (\ref{6}) in the case, when $S_{j}^{z}\geq 0$%
.

In order to exclude stiff equations, for $S_{j}^{z}<0$ it is convenient to
introduce other variables $Q$ and $P$ as follows

\[
b=\frac{Q+iP}{\sqrt{2}},\qquad b^{\ast }=\frac{Q-iP}{\sqrt{2}} 
\]
The equations of motion for $Q$ and $P$ have the form:

\begin{eqnarray}
\frac{dQ}{d\tau } &=&\frac{\partial {\cal H}/\hbar }{\partial {\bf S}}\cdot 
\frac{\partial {\bf S}}{\partial P}=\gamma {\bf H}_{{\rm eff}}\cdot \frac{%
\partial {\bf S}}{\partial P},  \nonumber \\
\frac{dP}{d\tau } &=&-\frac{\partial {\cal H}/\hbar }{\partial {\bf S}}\cdot 
\frac{\partial {\bf S}}{\partial Q}=-\gamma {\bf H}_{{\rm eff}}\cdot \frac{%
\partial {\bf S}}{\partial Q}.  \label{eq-QP}
\end{eqnarray}

\section{Energy transfer}

Mathematically the question of energy transfer between ${\cal H}_{{\rm Z}}$, 
${\cal H}_{{\rm exch}}$, ${\cal H}_{{\rm anis}}$ and ${\cal H}_{{\rm dd}}$.
is associated with the Eq.\ (\ref{hameq}) with ${\cal O}={\cal H}_{{\rm Z}}$%
: 
\begin{equation}
i\hbar \,\frac{d{\cal H}_{{\rm Z}}}{dt}=[\![\,{\cal H}_{{\rm Z}},{\cal H}_{%
{\rm Z}}+{\cal H}_{{\rm exch}}+{\cal H}_{{\rm anis}}+{\cal H}_{{\rm dd}%
}\,]\!]  \label{B1}
\end{equation}
It is obvious that $[\![\,{\cal H}_{{\rm Z}},{\cal H}_{{\rm Z}}\,]\!]=0$.
Direct energy transfer from the Zeeman energy to the exchange energy is
always forbidden, because 
\begin{equation}
\lbrack \![\,{\cal H}_{{\rm Z}},{\cal H}_{{\rm exch}}\,]\!]=0.  \label{B2}
\end{equation}

The commutator $\,[\![\,{\cal H}_{{\rm Z}},{\cal H}_{{\rm anis}}\,]\!]\,$
depends on the orientation of the external magnetic field. If the external
magnetic field is parallel to $z$ axis (${\bf H}=(0,0,H)$) then this
commutator is equal to zero. However, even a slight deviation of ${\bf H}$
from the $z$ axis permits a possibility to transform in sequence the Zeeman
energy to the anisotropy energy and the latter to the exchange energy as
long as 
\begin{equation}
\lbrack \![\,{\cal H}_{{\rm exch}},{\cal H}_{{\rm anis}}\,]\!]\neq 0.
\label{B3}
\end{equation}

\newpage

Captions

Fig.1.

Model of micromagnetic grain ($4\times 4\times 4$ subgrains).

Fig.2.

New coordinates.

Fig.3.

Short (a) and long (b) time evolution of magnetization in the case of very
small deviation of reversal magnetic field from the anisotropy axis.

Fig.4.

Short (a) and long (b) time evolution of magnetization in the case of
moderate deviation of reversal magnetic field from the anisotropy axis.

Fig.5.

Transformations of Zeeman, anisotropy and exchange energies for short (a,c)
and long (b,d) time intervals. The dipole interactions are neglected in
(a,b) and are taken into account in (c,d). $H_{z}=-0.79H_{{\rm K}}$ is the
onset of reversal, $H_{x}/H_{z}=0.05$, $\delta _{{\rm w}}/\delta _{{\rm ex}%
}=1.08$.

Fig.6.

Time evolution of magnetization for various reversal fields at $%
H_{x}/H_{z}=0.05$, $\delta _{{\rm w}}/\delta _{{\rm ex}}=1.08$ (dipole
interactions are neglected).

Fig.7.

Average asymptotic longitudinal component and absolute value of
magnetization versus applied reverse field. Conditions correspond to Fig.6.

Fig.8

Time evolution of magnetization for various grain sizes $\delta _{{\rm s}%
}/\delta _{{\rm w}}$. $H_{z}/H_{{\rm K}}=-0.79$, $H_{x}/H_{z}=0.05$, $\delta
_{{\rm w}}/\delta _{{\rm ex}}=1.08$ (dipole interactions are neglected).

Fig.9.

Short (a) and long (b) time evolution of magnetization in the case of very
small deviation of reversal magnetic field from the anisotropy axis. The
strength of reversal field corresponds to the onset of reversal (dipole
interactions are taken into account).

Fig.10.

Short (a) and long (b) time evolution of magnetization in the case of
moderate deviation of reversal magnetic field from the anisotropy axis. The
strength of reversal field corresponds to the onset of reversal (dipole
interactions are taken into account).

Fig.11.

Time evolution of magnetization for various reversal fields at $%
H_{x}/H_{z}=0.05$, $\delta _{{\rm w}}/\delta _{{\rm ex}}=1.08$ (dipole
interactions are taken into account).

Fig.12.

Relative change in Zeeman energy versus applied reversal magnetic field.

Fig.13. Time evolution of magnetization for various grain sizes $\delta _{%
{\rm s}}/\delta _{{\rm w}}$. $H_{z}/H_{{\rm K}}=-0.79$, $H_{x}/H_{z}=0.05$, $%
\delta _{{\rm w}}/\delta _{{\rm ex}}=1.08$ (dipole interactions are taken
into account).

\end{document}